\documentclass[makeidx]{crmp-l}

\makeindex

\newcommand{\ZZ}{{\mathbb Z}}
\newcommand{\RR}{{\mathbb R}}
\newcommand{\NN}{{\mathbb N}}
\newcommand{\TT}{{\mathbb T}}
\newcommand{\CC}{{\mathbb C}}

\newtheorem{theorem}{Theorem}[section]
\newtheorem{lemma}[theorem]{Lemma}
\newtheorem{proposition}[theorem]{Proposition}
\theoremstyle{definition}
\newtheorem{definition}[theorem]{Definition}
\newtheorem{example}[theorem]{Example}

\theoremstyle{remark}
\newtheorem{remark}[theorem]{Remark}

\numberwithin{equation}{section}

\begin{document}
\sloppy
\title[Spectral theory of one-dimensional quasicrystals]
{Gordon-type arguments in the spectral theory of one-dimensional quasicrystals}
\author{David Damanik}
\address{Fachbereich Mathematik, Johann Wolfgang Goethe-Universit\"at, 
D-60054 {}Frankfurt, Germany}
\curraddr{Department of Mathematics 253-37, California Institute of Technology, 
Pasadena, CA 91125, USA}
\email{damanik@its.caltech.edu}
\thanks{The author was supported by the German Academic Exchange Service 
(DAAD) through Hochschulsonderprogramm III (Postdoktoranden).}

\begin{abstract}
We review the recent developments in the spectral theory of discrete 
one-dimensional Schr\"odinger operators with potentials generated by 
substitutions and circle maps. We discuss how occurrences of local repetitive 
structures allow for estimates of generalized eigenfunctions. Among the recent 
applications of this general approach are almost sure and uniform results on 
the absence of eigenvalues as well as continuity of the spectral measures with 
respect to Hausdorff measures.
\end{abstract}

\maketitle

\section*{Introduction}
In 1983, two groups independently proposed and investigated a simple model, a 
discrete one-dimensional Schr\"odinger operator with a potential taking only 
two values, which was shown to exhibit unexpected and spectacular behavior. 
Kohmoto et al.~\cite{kkt} and Ostlund et al.~\cite{oprss} employed dynamical 
systems methods to study the scaling properties that were embodied in their 
model problem. It was argued that the eigenfunctions display a critical 
behavior in that they are neither extended nor localized. Moreover, this 
critical behavior seemed to be a universal feature of the model being quite 
independent of a modulation of the strength of the potential values. This was 
in sharp contrast to another popular model, the Harper operator, which exhibits 
such critical eigenfunctions only for some fixed modulation which in fact 
represents a sharp transition from extended to localized states.

One year later, in 1984, Shechtman et al.~published their discovery of a 
structure which has long range order without being globally translation 
invariant \cite{sbgc}. This discovery was essentially the birth of a whole new 
field, the investigation of quasicrystals, structures whose existence has been 
absolutely unexpected up to then, and it triggered intensive research 
activities including a reconsideration of the nature and definition of order, 
compare, for example, \cite{b3}.

{}Further structures with this property were soon discovered and on the 
theoretical side, models were proposed which reflect the observed phenomena. 
Chief among these were quasiperiodic structues that are constructed by a 
cut-and-project mechanism, which basically projects an ordered structure from 
a higher-dimensional space to the physical space. In 1986, Luck and Petritis 
showed in \cite{lp} that the model proposed by Kohmoto et al.~and Ostlund et 
al.~is naturally associated to such a cut-and-project structure. The features 
of this model were then generalized to related models by Kalugin et 
al.~\cite{kkl2} and Gumbs and Ali \cite{ga}. All these works concluded the 
critical behavior of eigenfunctions as well as the nowhere dense structure of 
the set of allowed energies which was in fact claimed to be a set of zero 
Lebesgue measure (compare in particular \cite{ko}). Also in 1986, using 
symbolic dynamics methods, Casdagli presented a fine anaysis of the 
{}Fibonacci dynamical system which further supported this claim \cite{c}.

In a 1987 paper, S\"ut\H{o} pursued a rigorous study of these phenomena for 
the basic original model, the {\it {}Fibonacci Hamiltonian} \cite{s5}. He was 
able to confirm parts of the observations, namely delocalized states and thus 
absence of point spectrum, and, for sufficiently large modulation of the 
potential, nowhere dense structure of the spectrum. He complemented these 
results in 1989 by showing that the spectrum is indeed always a set of measure 
zero \cite{s6}. Absence of absolutely continuous spectrum then follows 
immediately. In the same year, Bellissard et al.\ obtained these results 
for a more general class by essentially the same strategy \cite{bist}. 
It was thus rigorously established that those operators exhibit purely 
singular continuous zero-measure spectrum, reflecting the fact that they 
model one-dimensional structures being intermediate between periodic (leading 
to absolutely continuous spectrum) and disordered (leading to pure point 
spectrum).

{}From a mathematical point of view, the occurrence of purely singular 
continuous spectrum, however, was still considered to be a curiosity joining 
the ``constructed'' toy model examples by Pearson \cite{p2}. Only in the 
mid-90's did Simon and co-workers aim at a conceptual understanding of 
singular continuous spectral measures. In a series of papers 
\cite{s2,dms,js,djls,ss,s3,s4}, they not only exhibited many new examples of 
operators with this spectral type, they even showed that the occurrence of 
purely singular continuous spectrum is generic in an appropriate sense.

Moreover, in the course of this decade, building upon the landmark papers 
mentioned above, the occurrence of purely singular continuous spectrum has 
been shown to be a universal feature of operators associated to structures 
displaying aperiodicity at an intermediate level. To review these developments 
is the purpose of the present article. In 1994, S\"ut\H{o} contributed a review in similar spirit to a Les Houches winter school \cite{s7}. We therefore put particular emphasis on ideas and approaches that were introduced since then. Moreover, we shall center these approaches around one core idea which is due to Gordon and dates back to  
the 70's \cite{g2}. It was recently realized that the philosophy embodied in 
Gordon's paper may serve as a universal tool in the theory as the necessary 
input, local repetitive structures, is always present in the proposed models. 
{}For further background and an introduction to the relevant basics in operator 
theory, we refer the reader to S\"ut\H{o}'s article and, as further 
introductory reading, we want to mention \cite{bg3,d1}.

Our main goal here is to explain what quantities to look at when studying 
spectral properties of one-dimensional quasicrystal models, to present 
strategies and methods of their investigation which either have been 
successfully applied in the past or which seem promising when tackling some 
of the open problems, and to list some of these problems which appear to be 
important and interesting.

The organization is as follows. We start by presenting, in Section~1, 
essential parts of the theory of ergodic families of discrete one-dimensional 
Schr\"odinger operators which provides a very useful framework
to work within due to strong general results and proof strategies. Section~2 
presents two classes of such families which we will focus on, both being 
natural generalizations of the family of {}Fibonacci operators. It is shown how 
these classes fit into the framework, and known results for these families are 
recalled. The fundamental Gordon idea, upon which the core message of this 
paper is based, is discussed in Section~3. A certain variant of the Gordon 
method suggests investigating traces of some unimodular $2 \times 2$-matrices 
associated to an operator. Section~4 explains how useful bounds on these traces 
can be obtained by studying a dynamical system which is induced by hierarchical
structures in the potential value arrangement. A recently introduced general 
strategy for obtaining results that hold for all members of a family of 
operators is presented in Section~5. It basically emphasizes a combinatorial 
point of view as opposed to measure theoretical type of arguments suggested by 
ergodic theory. Using this method, one may investigate quantities that do not 
behave nicely under very weak perturbations, such as the point spectrum. 
Consequently, Section~6 deals with results on the absence of point spectrum. 
We present three types of results which are of increasing completeness, the 
most complete, of course, being uniform results. The latter is shown to be 
accessible by combining the Gordon method, the bounds on the traces, and the 
combinatorial point of view. Parts of the results obtained in the proofs of 
absence of eigenvalues can in fact be used to show that the spectrum has 
Lebesgue measure zero as explained in Section~7. Section~8 is concerned with 
transport properties of one-dimensional quasicrystals and hence with the 
unitary groups generated by the operators. We discuss recent results and 
possible ways to obtain bounds on the time evolution. {}Finally, we present 
some open problems in Section~9.

\section{Ergodic families of Schr\"odinger operators}
In this section we recall the concept of ergodic families of Schr\"odinger 
operators which has proved to provide a convenient framework for the operators 
that are of interest in this article. {}For more detailed presentations we refer 
the reader to the books by Carmona-Lacroix \cite{cl} and Cycon et 
al.~\cite{cfks}. In particular, every application of the fundamental Kotani 
theory requires this framework, and this theory is at the heart of all results 
concerning absence of absolutely continuous spectrum and the zero-measure 
property for the operators under consideration. Our presentation does not 
strive for greatest generality but rather for an appropriate notion of ergodic 
family which comprises all the examples we want to discuss here.  

\begin{definition} 
Let $\Omega$ be a compact metric space and let $T:\Omega \rightarrow \Omega$ be 
a homeomorphism. The pair $(\Omega,T)$ is called a {\it topological dynamical 
system}\index{topological dynamical system}. Given some $\omega \in \Omega$, 
the set $\{T^n \omega : n \in \ZZ\}$ is called the {\it orbit}\index{orbit} of 
$\omega$. Denote by $\mathcal{B}$ the Borel $\sigma$-algebra of $\Omega$. A 
Borel probability measure $\mu$ is called {\it stationary}
\index{stationary measure} if 
$\mu(T(B)) = \mu(B)$ for every $B \in \mathcal{B}$. A Borel set $B$ is 
called {\it shift invariant}\index{invariant set} if $T(B)=B$. A stationary 
measure is called {\it ergodic}\index{ergodic measure} if any shift invariant 
set has measure zero or one. The topological dynamical system $(\Omega,T)$ is 
called {\it minimal}\index{minimal} if the orbit of every $\omega \in \Omega$ 
is dense. It is called {\it uniquely ergodic}\index{uniquely ergodic} if 
there exists a unique ergodic measure, and it is called 
{\it stricly ergodic}\index{stricly ergodic} if it is both minimal and 
uniquely ergodic.
\end{definition}
It is in fact well known that if there is a unique stationary measure $\mu$, 
then $\mu$ is necessarily ergodic \cite{w}. The two examples below will be of 
major importance in what follows.

\begin{example}\label{ex1}
Let $\Omega=\TT \simeq [0,1)$ and $T \omega = \omega + \alpha \mod 1$ for some 
irrational $\alpha \in (0,1)$. It is well known that the Lebesgue measure on 
$\TT$ is the unique stationary measure (i.e., the system is uniquely ergodic) 
and that every orbit is dense (i.e., the system is minimal). Hence, 
$(\Omega,T)$ is strictly ergodic.
\end{example}
Before turning to the next example let us introduce some notation. 

\begin{definition}
Let $\mathcal{A}= \{a^{}_1,\ldots,a_s\}$ be a finite set, called 
{\it alphabet}\index{alphabet}. Endow $\mathcal{A}$ with the discrete topology.
The $a_i$ are called {\it symbols}\index{symbol} or {\it letters}
\index{letter}, the elements of 
$\mathcal{A}^* = \bigcup_{k \ge 1} \mathcal{A}^k$ are called {\it words}
\index{word}. We denote by $|v|$
the {\it length}\index{length} of a word $v \in \mathcal{A}^*$ 
(i.e., $|v|=l$ if $v \in \mathcal{A}^l$). {}For $v,w \in \mathcal{A}^*$, 
$\#_v (w)$ denotes the number of occurrences of $v$ in $w$ 
(e.g., $\#_{aa} (aabaaa) = 3$). Let $\mathcal{A}^\NN, \mathcal{A}^\ZZ$ 
denote the sets of one-sided and two-sided infinite sequences, called 
{\it infinite words}\index{infinite word}, over $\mathcal{A}$, both being 
equipped with product topology which is easily seen to be a metric topology. 
Given a finite or infinite word $w$, a finite word $v$ is called a 
{\it subword}\index{subword} or {\it factor}\index{factor} of $w$ if there 
are (finite or infinite) words $r,s$ such that $w=rvs$, with the obvious 
definition of concatenation of words. We define 
$F_w = \{y:y$ is a factor of $w\}$ and $F_w(n) = {}F_w \cap \mathcal{A}^n$, 
$n \in \NN$. The {\it complexity function}\index{complexity function} 
$p_w:\NN \rightarrow \NN_0$ corresponding to $w$ is given by 
$p_w(n) = |F_w(n)|$, $|\cdot|$ denoting cardinality.
\end{definition}

\begin{example}\label{ex2}
Let $\mathcal{A}$ be an alphabet. The shift $T$ on $\mathcal{A}^\ZZ$ is
defined by $(T \psi )_n = \psi_{n+1}$ for $\psi \in \mathcal{A}^\ZZ$. Let 
$\Omega \subseteq \mathcal{A}^\ZZ$ be closed and invariant under $T$, that is, 
$T \Omega = \Omega$. The topological dynamical system $(\Omega,T)$ is called a 
{\it subshift}\index{subshift}. Let us consider subshifts that are generated 
as follows. Given some $\psi \in \mathcal{A}^\NN$, we define $\Omega_\psi$ to 
be the set of two-sided sequences having all their subwords occur in $\psi$, 
that is,
$$
   \Omega_\psi \; = \; 
   \{ \omega \in \mathcal{A}^\ZZ : {}F_\omega \subseteq {}F_\psi \}\, .
$$
It is clear that $\Omega_\psi$ is closed and invariant. Moreover, unique 
ergodicity and strict ergodicity of $(\Omega_\psi,T)$ can be characterized in 
terms of frequencies of subwords as follows (cf. \cite{q}). The subshift is 
uniquely ergodic if and only if for every $v \in {}F_\psi$ there is a number 
$d_\psi(v) \ge 0$, the \textit{frequency of $v$ in $\psi$}\index{frequency}, 
such that for every $k \in \NN$, we have 
$$
  \frac{\#_v(\psi_k \ldots \psi_{k+n-1})}{n} 
       \; \longrightarrow \; d_\psi(v)
$$
as $n \rightarrow \infty$, uniformly in $k$. Minimality of $(\Omega,T)$ is 
equivalent to the fact that every word in $F_\psi$ occurs infinitely often in 
$\psi$ and the gaps between consecutive occurrences are bounded by a constant 
depending on the word, that is, the occurrences of every word are relatively 
dense. Thus, the subshift is strictly ergodic if and only if it is uniquely 
ergodic with strictly positive frequencies, that is, $d_\psi(v) > 0$ for all 
$v \in {}F_\psi$. In the uniquely ergodic case, the unique stationary measure 
$\mu$ obeys 
\begin{equation}\label{cylmeasfreq}
   \mu ( \{ \omega \in \Omega : \omega_m \ldots \omega_{m + |v| - 1} = v \} ) 
   \; = \; d_\psi(v)
\end{equation}
for every $v \in {}F_\psi$ and every $m \in \ZZ$. That is, the measure of some 
cylinder set can be determined by studying the frequency of the defining word. 
\end{example}

\begin{definition}
Given a topological dynamical system $(\Omega,T)$, an ergodic measure $\mu$, 
and a measurable function $g:\Omega \rightarrow \RR$, one defines for each 
$\omega \in \Omega$ a two-sided infinite sequence 
$V_\omega :\ZZ \rightarrow \RR$ by $V_\omega(n) = g(T^n \omega)$. 
This gives rise to a discrete one-dimensional Schr\"odinger operator 
$H_\omega$ on $\ell^2(\ZZ)$ which acts on some $\phi \in \ell^2(\ZZ)$ by 
$$
  (H_\omega \phi) (n) \; = \; \phi(n+1) + \phi(n-1) + V_\omega(n) \phi(n) \, .
$$
The family $(H_\omega)_{\omega \in \Omega}$ is called an {\it ergodic family 
of Schr\"odinger operators}\index{ergodic family of Schr\"odinger operators}. 
\end{definition}
The striking fundamental result, which motivates the choice of this framework 
even for deterministic models such as the ones we consider in this paper, is 
the following result which essentially says that the spectrum and the spectral 
type are deterministic up to sets of measure zero.

\begin{theorem}[Pastur, Kunz-Souillard]\label{basic}
Let $(H_\omega)_{\omega \in \Omega}$ be an ergodic family of Schr\"odinger 
operators. Then there exist sets $\Omega_0 \subseteq \Omega$, 
$\Sigma, \Sigma_{{\rm pp}}, \Sigma_{{\rm sc}}, \Sigma_{{\rm ac}} \subseteq \RR$ 
such that $\mu(\Omega_0)=1$ and $\sigma(H_\omega) = \Sigma, 
\sigma_{{\rm pp}}(H_\omega) = \Sigma_{{\rm pp}},
\sigma_{{\rm sc}}(H_\omega) = \Sigma_{{\rm sc}},
\sigma_{{\rm ac}}(H_\omega) = \Sigma_{{\rm ac}}$ for every 
$\omega \in \Omega_0$.
\end{theorem}
A proof of this result can be found in \cite{p,ks}. However, for a discussion 
of this result and most of what follows in this section, the reader could also 
consult the books \cite{cl,cfks}. In fact, an additional assumption even allows 
for a strengthening of Theorem \ref{basic}.

\begin{definition}
An ergodic family of Schr\"odinger operators $(H_\omega)_{\omega \in \Omega}$ 
is called {\it minimal} if for each pair $\omega_1,\omega_2 \in \Omega$, the 
sequence $V_{\omega^{}_1}$ is a pointwise limit of translates of 
$V_{\omega^{}_2}$.
\end{definition}

\begin{remark}\label{minsysminops}
Note that if the family is generated by a dynamical system as defined in 
Example \ref{ex2} along with $g(\omega)=f(\omega_0)$ where 
$f:\mathcal{A} \rightarrow \RR$ is arbitrary, then minimality of the family 
follows from minimality of the dynamical system $(\Omega,T)$.
\end{remark}

\begin{theorem}\label{minunif}
Let $(H_\omega)_{\omega \in \Omega}$ be a minimal ergodic family of 
Schr\"odinger operators. Then, we have $\sigma(H_\omega) = \Sigma, 
\sigma_{{\rm ac}}(H_\omega) = \Sigma_{{\rm ac}}$ for every $\omega \in \Omega$.
\end{theorem}
The statement about the spectrum is already part of the folklore and is
essentially contained in \cite{rs1}. The result on the absolutely continuous
spectrum was recently obtained by Last and Simon \cite{ls}. Thus, given a 
minimal ergodic family, one can pick any member of the family when studying 
the spectrum or the absolutely continuous spectrum. This is a clear motivation 
for embedding even a deterministic model into this framework since it may well 
be that another member of the family, not the one we started with, is easier 
to study. Let us remark that minimality does not imply constancy of 
$\sigma_{{\rm pp}}(H_\omega)$ or $\sigma_{{\rm sc}}(H_\omega)$ as there are 
explicit counterexamples \cite{js}.

Let us now turn to a beautiful theory which has been termed 
{\it Kotani theory}\index{Kotani theory}. The results we shall describe below indeed 
form the core of much of the theory of ergodic families of Schr\"odinger operators in 
one dimension and certainly provide a basis for most results on {}Fibonacci-type 
operators which have been obtained so far. Given a family 
$(H_\omega)_{\omega \in \Omega}$, it is often very useful to consider the 
associated eigenvalue equation in difference sense, that is,
\begin{equation}\label{eve}
   \phi(n+1) + \phi(n-1) + V_\omega(n) \phi(n) \; = \; E \phi(n)\, ,
\end{equation}
where $E \in \CC$ and $\phi$ is just required to be a two-sided sequence,
$\phi:\ZZ \rightarrow \CC$. Some of the most useful tools in one-dimensional 
Schr\"odinger operator theory are results that establish a link between the 
behavior of solutions to (\ref{eve}) and spectral properties of the operator 
since the former are to some extent relatively easy to investigate. In fact, 
Kotani theory provides a link of this kind in the case of the ergodic 
framework. Connections in the deterministic case have been found by Gilbert, 
Pearson, and Khan \cite{gp,g1,kp} (see Jitomirskaya-Last \cite{jl1,jl2,jl3} 
for an extension of the results and a simplification of the proof) and by Last 
and Simon \cite{ls}. Let us recall a standard reformulation of (\ref{eve}). 
Given a two-sided sequence $\phi$ we define $\Phi: \ZZ \rightarrow \CC^2$ by
$$
  \Phi(n) \; = \; 
  \left( \begin{array}{c} \phi(n+1) \\ \phi(n) \end{array} \right) \, .
$$
Defining
\begin{align*}
T_{E,\omega}(n) & = \left( \begin{array}{cc} E - V_\omega(n) & -1 \\ 
                                  1 & 0 \end{array} \right),\\
M_{E,\omega}(n) & = \left\{ \begin{array}{ll} 
                     T_{E,\omega}(n) \times \cdots \times T_{E,\omega}(1), 
                     & n \ge 1, \\ I, & n=0, \\ 
                    T_{E,\omega}(n+1)^{-1} 
          \times \cdots \times T_{E,\omega}(0)^{-1}, & n \le -1, 
         \end{array}\right. 
\end{align*}
one may easily check that 
\begin{center}
  $\phi$ solves (\ref{eve}) $\;\Leftrightarrow\;$ $\Phi$ solves 
  $\Phi(n) = M_{E,\omega}(n)\Phi(0)$ for every $n \in \ZZ$.
\end{center}
The matrices $M_{E,\omega}(\cdot)$ are called {\it transfer matrices}
\index{transfer matrix}. They have determinant $1$ since they are products of 
the {\it elementary transfer matrices} $T_{E,\omega}(\cdot)$ which obviously 
have determinant $1$. The linear space of solutions to (\ref{eve}) for fixed 
$E$ is two-dimensional, as can be seen from the above relation. Consider the 
two solutions $\phi_{1,2}$ induced by the initial conditions 
$\phi_1(0) = \phi_2(1) = 0$, $\phi_1(1) = \phi_2(0) = 1$. Then we also have 
$$
   M_{E,\omega}(n) \; = \; 
  \left( \begin{array}{cc} \phi_1(n+1) & \phi_2(n+1) \\ 
  \phi_1(n) & \phi_2(n) \end{array} \right).
$$
Thus, the matrix contains all information about $\phi_{1,2}$ and hence all 
solutions in a pointwise way. In particular, bounds on $\|M_{E,\omega}(n)\|$, 
for example, yield bounds on $\|\Phi(n)\|$ for all solutions. It turns out to 
be useful to distinguish between exponential and sub-exponential growth of 
$\|M_{E,\omega}(n)\|$. Thus, we let
$$
   \gamma_{\omega,\pm}(E) \; = \; 
   \lim_{n \rightarrow \pm \infty} \frac{1}{n} \ln \|M_{E,\omega}(n)\|\, ,
$$
provided the limit exists. Regarding existence of this limit, the following 
has been obtained in \cite{fk}.

\begin{theorem}[Furstenberg-Kesten]
{}For every $E \in \CC$, there exist $\Omega_E \subseteq \Omega$ and 
$\gamma(E)  \in \RR$ such that $\mu(\Omega_E) = 1$ and for every 
$\omega \in \Omega_E$, $\gamma_{\omega,\pm}(E)$ exist and are equal to 
$\gamma(E)$, that is, $\gamma_{\omega,+}(E)=\gamma_{\omega,-}(E)=\gamma(E)$.
\end{theorem}
The number $\gamma(E)$ is called the {\it Lyapunov exponent}
\index{Lyapunov exponent}. 

\begin{theorem}[Osceledec]\label{osc}
Suppose that for some $E \in \CC$, $\gamma(E) > 0$. Then, for every 
$\omega \in \Omega_E$, there exist solutions $\phi_d^+,\phi_d^-$ of 
$H_\omega \phi = E \phi$ such that $\phi_d^\pm$ decays exponentially at 
$\pm \infty$, respectively, at the rate $-\gamma(E)$. Moreover, every solution 
which is linearly independent of $\phi_d^+$ {\rm (}resp., $\phi_d^-${\rm )} 
grows exponentially at $+ \infty$ {\rm (}resp., $- \infty${\rm )} at the rate 
$\gamma(E)$.
\end{theorem}
See \cite{ls,o,r2}. Thus, in the case of a positive Lyapunov exponent, one has 
a complete understanding of the asymptotics of the solutions at infinity. 

Kotani theory now establishes a link between the Lyapunov exponent and the 
absolutely continuous spectrum. Define
$$
   A \; = \; \{ E \in \RR : \gamma(E) = 0\} \, .
$$
The {\it essential closure} $\overline{S}^{{\rm ess}}$ of a set 
$S \subseteq \RR$ is defined by
$$
   \overline{S}^{{\rm ess}} \; = \; 
  \{ E \in \RR : |(E-\varepsilon , E+\varepsilon ) \cap S| > 0 
       \;\; \forall \varepsilon > 0 \} \, ,
$$
where $|\cdot|$ denotes Lebesgue measure. In particular, 
$\overline{S}^{{\rm ess}} = \emptyset$ for every set $S$ of zero Lebesgue 
measure.

\begin{theorem}[Ishii-Pastur-Kotani]
$\Sigma_{{\rm ac}} = \overline{A}^{{\rm ess}}$.
\end{theorem}
{}For a proof of the inclusion ``$\subseteq$'' the reader may consult 
\cite{i,p} (see also \cite{b} for an alternative proof using Gilbert-Pearson
theory), the opposite inclusion has been treated in \cite{k2} (see \cite{s1} 
for an adaptation to the discrete case). The following corollary, obtained in 
\cite{k3}, to the proof given in \cite{k2,s1} is of great interest to us since 
all the potentials we shall be dealing with take only finitely many values. 
Moreover, the additional assumption of aperiodicity is rather non-restrictive 
since there is a well-established theory treating the case of periodic 
potentials; see, for example, \cite{rs4}. 

\begin{theorem}[Kotani]\label{efaprop}
If the potentials $V_\omega$ are aperiodic and take only finitely many values, 
then $|A|=0$. In particular, $\Sigma_{{\rm ac}}=\emptyset$.
\end{theorem}

\section{Models generated by circle maps and primitive substitutions}
In this section we present the two classes of ergodic families we shall 
discuss in the sequel. Both classes are natural extensions of different 
aspects of the {}Fibonacci model, one generalizing its quasiperiodicity 
(models generated by circle maps) or, when restricted to a subclass, its 
word complexity properties (models generated by Sturmian sequences), the 
other one generalizing its self-similar structure (models generated by 
primitive substitutions). We show how they fit into the general framework 
presented in the preceding section and recall known results for these classes. 
{}For the sake of brevity we introduce the following notions.

\begin{definition}
A family $(H_\omega)_{\omega \in \Omega}$ is called {\it EFA} if it is an 
ergodic family of Schr\"odinger operators such that the potentials $V_\omega$ 
are aperiodic and take only finitely many values. It is called {\it MEFA} if 
it is EFA and minimal.
\end{definition}
With this convention at hand we infer from the above results that EFA (MEFA) 
families exhibit almost surely (uniformly) purely singular spectrum, that is, 
$\Sigma = \Sigma_{{\rm pp}} \cup \Sigma_{{\rm sc}}$. More precisely, the set 
$A$ associated to such a family has zero Lebesgue measure. {}Furthermore, MEFA 
families have constant spectrum. In fact, all the classes of families 
$(H_\omega)_{\omega \in \Omega}$ we shall present and discuss now will be 
MEFA families, so let us bear in mind that when studying the spectral type, 
we only need to distinguish between point spectrum and singular continuous 
spectrum since all these families have empty absolutely continuous spectrum 
for all $H_\omega$, $\omega \in \Omega$; thus the latter does not present any 
issue at all.\\[5mm]
{\bf Circle map models}: A circle map model is parametrized by three
parameters, namely, an irrational {\it rotation number}\index{rotation number} 
$\alpha \in (0,1)$, an {\it interval length}\index{interval length} 
$\beta \in (0,1)$, and a {\it coupling constant}\index{coupling constant} 
$\lambda \in \RR \backslash \{0\}$. There are two ways to choose $\Omega,T,g$, 
both of which have been used in the past. Their mutual relation is, for 
example, given in \cite{dl3}. The first way follows Example \ref{ex1}. Thus, 
let $\Omega = \TT \simeq [0,1)$ and 
$T:\Omega \rightarrow \Omega,\; \omega \mapsto \omega + \alpha \mod 1$. 
As noted above, $(\Omega,T)$ is strictly ergodic with the Lebesgue measure 
on $\TT$ as unique ergodic measure $\mu$. Let $g$ be given by 
$g(\omega) = \lambda \cdot \chi^{}_{[1-\beta,1)}(\omega)$. This yields 
potentials $V_\omega (n) = \lambda \cdot 
\chi^{}_{[1-\beta,1)}(\alpha n+\omega \mod 1)$. 
The other possibility of associating a family of potentials starts with the 
one-sided sequence 
$v_{\alpha,\beta,\theta}(n)=\chi^{}_{[1-\beta,1)}(n\alpha + \theta \mod 1)$, 
$n \in \NN$, and follows the lines of Example \ref{ex2}. Define 
$\Omega_{\alpha,\beta} = \Omega_{v_{\alpha,\beta,\theta}}$. The notation is 
justified since the subshift does not depend on $\theta$. It was shown by Hof 
in \cite{h} that the dynamical system $(\Omega_{\alpha,\beta},T)$ is strictly 
ergodic. The function $g$ generating the potentials is in this case given by 
$g(\omega) = f(\omega_0)$, where $f(0)=0$, $f(1)=\lambda$. In the case 
$\alpha=\beta$ we write $v_{\alpha,\theta}$ instead of 
$v_{\alpha,\alpha,\theta}$ and $\Omega_\alpha$ instead of 
$\Omega_{\alpha,\alpha}$, and we call the dynamical system 
$(\Omega_\alpha,T)$ as well as the resulting potentials 
{\it Sturmian}\index{Sturmian potential}. Sturmian potentials have an interesting 
combinatorial property. Consider a one-sided sequence $s$. Recall that the 
complexity function $p_s(n)$ counts the number of factors of length $n$ in $s$. 
One can show (see \cite{loth1,loth2} for this and much more on combinatorics 
on words in general and Sturmian sequences in particular, compare also 
\cite{lp2}) that $p_s(n) \le n$ for some $n$ implies that $s$ is ultimately 
periodic. Thus, any non-ultimately periodic sequence $s$ obeys 
$p_s(n) \ge n+1$ for every $n$. Sequences $s$ having $p_s(n) = n+1$ for all 
$n$ are called {\it Sturmian}\index{Sturmian sequence}. The terminology is now 
motivated by the fact that every $v_{\alpha,\theta}$ is a Sturmian sequence 
and every $\{0,1\}$-valued Sturmian sequence coincides with the restriction of 
some element of an appropriate $\Omega_\alpha$ to $\NN$.

Our discussion of this relation involving one-sided sequences may appear
somewhat awkward, but the corresponding relation for two-sided sequences is 
not true, that is, there exist $\{0,1\}$-valued two-sided sequences $s$ with 
complexity function obeying $p_s(n) = n+1$ which, however, do not belong to 
some $\Omega_\alpha$ (consider, e.g., the sequence $s$ defined by $s_n = 0$ 
for $n < 0$ and $s_n = 1$ for $n \ge 0$), compare \cite{ch} for a complete 
characterization.

Let us note the following concerning families generated by 
circle maps.

\begin{proposition}
Both ways of generating an operator family $(H_\omega)_{\omega \in \Omega}$ 
corresponding to the parameters $\alpha,\beta,\lambda$ induce MEFA families.
\end{proposition}
{}For fixed parameter values, the operator family induced by the dynamical 
system on the torus is contained in the operator family obtained from the 
subshift. Interestingly, the latter family is strictly larger even though both 
families are minimal \cite{dl3}! The Sturmian model generated by the golden 
mean $\alpha = \frac{\sqrt{5}-1}{2}$ is called the {\it {}Fibonacci model}
\index{Fibonacci operator}.\\[5mm]
{\bf Models generated by primitive substitutions}: Let $\mathcal{A}$ be an 
alphabet. A {\it substitution}\index{substitution} $S$ is a map 
$S : \mathcal{A} \rightarrow \mathcal{A}^*$. $S$ can be extended morphically 
to $\mathcal{A}^*$ (resp.,
$\mathcal{A}^{\NN}$) by $S(b_1 \ldots b_n) = S(b_1) \ldots S(b_n)$ 
(resp., $S(b_1 b_2 b_3\ldots) = S(b_1) S(b_2) S(b_3)\ldots$). $S$ is 
called {\it primitive}\index{primitive} if there exists $k \in \NN$ such 
that for every
$a \in \mathcal{A}$, $S^k(a)$ contains every symbol from $\mathcal{A}$. 
Prominent examples of primitive substitutions are given by
\begin{center}
\begin{tabular}{ll}
$a \mapsto ab$, $b \mapsto a$ & {}Fibonacci\index{substitution!Fibonacci},\\
$a \mapsto ab$, $b \mapsto aa$ & 
period doubling\index{substitution!period doubling},\\
$a \mapsto ab$, $b \mapsto aaa$ & 
binary non-Pisot\index{substitution!binary non-Pisot},\\
$a \mapsto ab$, $b \mapsto ba$ & 
Thue-Morse\index{substitution!Thue-Morse},\\
$a \mapsto ab$, $b \mapsto ac$, $c \mapsto db$, $d \mapsto dc$ &
Rudin-Shapiro\index{substitution!Rudin-Shapiro}. 
\end{tabular}
\end{center}
A fixed point $u \in \mathcal{A}^{\NN}$ of $S$ is called 
{\it substitution sequence}\index{substitution sequence}. The existence of 
such a fixed point is ensured by the following conditions,
 
\begin{itemize}
\item there exists a letter $a \in \mathcal{A}$ such that the first letter of 
$S(a)$ is $a$,
\item $\lim_{n \rightarrow \infty} |S^n(a)| = \infty$,
\end{itemize}
which are easily seen to hold for a suitable power of $S$ if $S$ is
primitive. Without loss of generality (since any power of $S$ is primitive if 
$S$ is primitive), we assume this power to be equal
to one. In this case, $u = \lim_{n \rightarrow \infty} S^n(a)$
exists and is a substitution sequence. Define $\Omega$ by 
$\Omega = \Omega_u$. If $S$ is primitive, $\Omega$ does not depend on the 
choice of the substitution sequence. The subshift $(\Omega,T)$ is called the 
{\it substitution dynamical system associated to $S$}
\index{substitution dynamical system} and it is stricly ergodic \cite{q}. 
Choose some function $f:\mathcal{A} \rightarrow \RR$ and define 
$g(\omega) = f(\omega_0)$.

\begin{proposition}
Suppose that the substitution $S$ is primitive, the substitution sequence 
$u$ is not ultimately periodic, and the function $f$ takes at least two values. 
Then the induced operator family $(H_\omega)_{\omega \in \Omega}$ is MEFA.
\end{proposition}
In case of the {}Fibonacci substitution the subshift is equivalent to the 
Sturmian subshift corresponding to $\alpha = \frac{\sqrt{5}-1}{2}$ via 
$a \mapsto 1, b \mapsto 0$. Thus the corresponding families of operators 
coincide (up to a spectral shift).\\[5mm]
These two classes of MEFA families have been studied extensively in the past;
see \cite{bist,bit,d2,dl4,dl1,dl2,dp1,it,irt,hks,j3,k1,r,s5,s6} 
and 
\cite{b2,bbg1,bbg2,bg1,d3,d4,d6,dp2,h,hks} for 
some important contributions 
in the case of circle map models and substitution models, respectively. 
The results comprise in particular singular continuity of spectral measures, 
zero Lebesgue measure of the spectrum, Gap labelling via $K$-theory, opening 
of the gaps at low coupling, continuity of gap boundaries with respect to the 
rotation number, and uniform existence of the Lyapunov exponent for large 
subclasses.

\section{Pointwise methods and variants of the Gordon criterion}
This section is concerned with methods in the spectral theory of some fixed 
Schr\"odinger operator with particular emphasis on several variants of an 
idea originally due to Gordon \cite{g2}. The methods we present can be applied, 
for example, to a fixed member of some ergodic family 
$(H_\omega)_{\omega \in \Omega}$. 

Consider a bounded function $V:\ZZ \rightarrow \RR$ and the associated 
Schr\"odinger operator
\begin{equation}\label{detop}
   (H\phi)(n) \; = \; \phi(n+1) + \phi(n-1) + V(n)\phi(n)
\end{equation}
along with the difference equation
\begin{equation}\label{deteve}
   \phi(n+1) + \phi(n-1) + V(n)\phi(n) \; = \; E \phi(n) \, ,
\end{equation}
where $E \in \CC$. Similarly to the above, we introduce a reformulation of
(\ref{deteve}) in terms of {\it transfer matrices}\index{transfer matrix} 
$M_E(n)$,

\begin{align*}
\Phi(n) & \; = \; \left( \begin{array}{c} \phi(n+1) \\ 
                  \phi(n) \end{array} \right),\\
T_E(n)  & \; = \; \left( \begin{array}{cc} E - V(n) & -1 \\ 
                  1 & 0 \end{array} \right),\\
M_E(n)  & \; = \; \left\{ \begin{array}{ll} T_E(n) \times \cdots 
                  \times T_E(1), & n \ge 1,\\ I, & n=0, \\ 
                  T_E(n+1)^{-1} \times \cdots \times T_E(0)^{-1}, 
                  & n \le -1. \end{array}\right. 
\end{align*}
Let us discuss ``Gordon-type arguments,'' 
that is, the exploitation of local 
repetitions. Recall that transfer matrices have determinant $1$, independently 
of the potential $V$, the energy $E$, and the site $n$. Thus, by the 
Cayley-Hamilton theorem\index{Cayley-Hamilton theorem}, the following 
universal equation holds,
\begin{equation}\label{char}
     M_E(n)^2 - {\rm tr}(M_E(n)) M_E(n) + I \; = \; 0 \, .
\end{equation}
Suppose now that $V$ repeats its values on the interval $\{1,\ldots,n\}$ once, 
that is,
\begin{equation}\label{once}
   V(j) \; = \; V(j + n), \; 1 \le j \le n \, .
\end{equation}
Due to the fact that the definition of transfer matrices is local, we infer 
that for any energy $E$, we have 
\begin{equation}\label{matsq}
    M_E(n)^2 \; = \; M_E(2n) \,.
\end{equation}
Plugging this into (\ref{char}), we obtain
\begin{equation}\label{charrep}
   M_E(2n) - {\rm tr}(M_E(n)) M_E(n) + I \; = \; 0 \, .
\end{equation}
Now consider any initial vector $\Phi(0)$, which we may assume to be 
normalized, that is, $\| \Phi(0) \| = 1$. We apply (\ref{charrep}) to 
$\Phi(0)$ and get
\begin{equation}\label{charrepapp}
    \Phi(2n) - {\rm tr}(M_E(n)) \Phi(n) + \Phi(0) \; = \; 0 \, .
\end{equation}
Since $\Phi(0)$ has norm $1$, either $\Phi(2n)$ or ${\rm tr}(M_E(n)) \Phi(n)$ 
has to have norm at least $\frac{1}{2}$. Thus,
\begin{equation}\label{lowernormbound}
  \max ( \|\Phi(n)\| , \|\Phi(2n)\| ) \; \ge \; 
  \frac{1}{2} \min \left( 1 , \frac{1}{|{\rm tr}(M_E(n))|} \right).
\end{equation}
If we can find, for some fixed energy $E$, a sequence $n_k 
\rightarrow \infty$ 
such that the potential repeats the values on $\{1,\ldots,n_k\}$ once and the 
sequence ${\rm tr}(M_E(n_k))$ remains bounded, then the right-hand side of 
(\ref{lowernormbound}) is strictly bounded away from zero on this sequence of 
sites for every initial vector! Thus, in this case no solution tends to zero 
and, in particular, $E$ is not an eigenvalue of $H$. Let us summarize this 
in the following lemma.

\begin{lemma}[two-block method\index{two-block method}]\label{zweiblock}
{}Fix a potential $V$ and an energy $E$. Suppose there is a sequence 
$n_k \rightarrow \infty$ and some $1 \le C < \infty$ such that we have for 
every $k$,
\begin{enumerate}
\item $V(j) = V(j + n_k)$, $1 \le j \le n_k$,
\item $|{\rm tr}(M_E(n_k))| \le C$.
\end{enumerate}
Then, $E$ is not an eigenvalue of $H$ and no solution of {\rm (\ref{deteve})}
tends to zero at $+\infty$. More precisely, for every $k$, every solution obeys
$$
    \max ( \|\Phi(n_k)\| , \|\Phi(2n_k)\| ) \;\ge\; \frac{1}{2C}\, .
$$
\end{lemma}

\begin{remark}\label{2bmodif}
Of course, there are obvious variations on this idea. {}First of all, it is not 
important that the squares are aligned at the origin; any other site will do. 
Similarly, the squares can also be aligned to their right side and one can work 
on the left half-line. In this case one gets that, once the modified conditions 
are satisfied, no solution tends to zero at $- \infty$ with similar uniform 
lower bound for the norms.
\end{remark}
But what if a study of transfer matrix traces is not feasible? The answer is 
simple, just find another block of repetition! The key ingredient in the above 
argument is the three-term expression in (\ref{charrep}) which, given largeness 
of one term, yields largeness of at least one of the others. Now suppose that 
we have a repetition of $V(1),\ldots,V(n)$ and that the trace of $M_E(n)$ is 
large. Even if we infer ``largeness'' of the middle term, this may only be due 
to the trace but not to the matrix (resp., the vector after application of the 
equation to an initial vector). This complication does not occur for the other 
terms. So, in case of a large trace, try to find a repetition of the potential 
values from $1$ to $n$ to the left and, if successful, apply (\ref{charrep}) to
$\Phi(-n)$ but retain normalization at the origin! This yields the equation
\begin{equation}\label{charrepappleft}
   \Phi(-n) - {\rm tr}(M_E(n)) \Phi(0) + \Phi(n) \; = \; 0 \, .
\end{equation}
Now, the middle term is large (a large factor times a normalized vector) and, 
again, this says that at least one of the other vectors has to be large. 
Quantitatively, we have the following lemma.

\begin{lemma}[three-block method\index{three-block method}]\label{dreiblock}
{}Fix a potential $V$. Suppose there is a sequence $n_k \rightarrow \infty$ such 
that we have for every $k$,
$$
    V(j - n_k) \; = \; V(j) = V(j + n_k) \; ,\quad 1 \le j \le n_k \, .
$$
Then for every energy $E$, we have that $E$ is not an eigenvalue of $H$ and 
no solution of {\rm (\ref{deteve})} tends to zero at both $\pm \infty$. More 
precisely, for every $k$, every solution obeys
$$
   \max ( \|\Phi(-n_k)\| , \|\Phi(n_k)\| , \|\Phi(2n_k)\| ) 
   \; \ge \; \frac{1}{2} \, .
$$
\end{lemma}

\begin{remark}\label{2badv}
In general it may well happen that for some energy $E$, every solution decays 
at either $+\infty$ or $-\infty$. This happens, for example, for energies 
outside the spectrum of $H$. Thus, even if one has cubes rather than only 
squares, the additional investigation of transfer matrix traces pays off in 
the form of a stronger conclusion.
\end{remark}
Lemma \ref{dreiblock} is very close to the original Gordon result \cite{g2}
(see also \cite{cfks}) which, however, requires another block of repetition.
It was stated and proved in this form by Delyon and Petritis \cite{dp1}. 
Lemma \ref{zweiblock} was proved by S\"ut\H{o} in \cite{s5}.

\section{Trace map characterization of the spectrum}
The two-block version of the Gordon criterion presented in Lemma 
\ref{zweiblock} suggests investigating both repetitive structures in the 
potential and traces of transfer matrices when trying to obtain bounds on 
solutions of the eigenvalue equation. In the case of Sturmian models or models 
generated by primitive substitutions, transfer matrix traces can be 
investigated by studying a (generalized) dynamical system, the 
{\it trace map}\index{trace map}, which 
is induced by hierarchical structures in the 
potentials. These are by definition present in potentials generated by 
substitutions and their presence in the Sturmian case can be exhibited 
using continued fraction expansion theory. The present section is concerned 
with a discussion of the correspondence between the dynamics of trace maps 
and the spectra of operators from the two classes. {}For further information 
on trace maps, we refer the reader to \cite{ap,bgj,br,kn,pww,r1,rb}.

The introduction of a trace map follows a universal program which can be 
summarized as follows. 
\begin{enumerate}
\item Exhibit a sequence of generating words obeying recursive relations.
\item Consider transfer matrices as being associated to finite words rather 
      than to infinite sequences from the subshift.
\item Translate the recursive relations to the level of transfer matrices.
\item Pass to the traces of these matrices using suitable identities for 
      unimodular $2 \times 2$-matrices.
\end{enumerate}
Among the models we are interested in, trace maps have been found for all 
Sturmian models and all substitution models. Let us indicate how to establish 
the above steps in these two cases.

Given some one-sided sequence $\psi$ such that the associated subshift 
$(\Omega_\psi,T)$ is minimal, it is easy to check that we have (this is 
essentially Gottschalk's theorem, see \cite{pet})
\begin{equation}\label{combihull}
   \Omega_\psi \; = \; \{ \omega \in \mathcal{A}^\ZZ : {}F_\omega = {}F_\psi \} \, .
\end{equation}
In the case of $\psi$ being a substitution sequence associated to some 
primitive substitution $S$, $\psi$ has the form 
$\psi = \lim_{n \rightarrow \infty} S^n(a)$ for a suitable $a \in \mathcal{A}$. 
In particular, we have
\begin{equation}\label{substhullword}
     \Omega_\psi \; = \; 
   \{ \omega \in \mathcal{A}^\ZZ : \forall w \in {}F_\omega \, 
  \exists n \in \NN \; {\rm such} \, {\rm that} \, w \in {}F_{S^n(a)} \}\, .
\end{equation}
Thus, the words $S^n(a)$ entirely determine the hull. Due to primitivity, any 
other sequence of the form $S^n(b)$, $b \in \mathcal{A}$, can be used. The set 
of words $S^n(b)$ where $n$ ranges over $\NN_0$ and $b$ ranges over 
$\mathcal{A}$ therefore serves as a good basis for a study of the local 
properties of the potential value arrangements. Moreover, among these words 
the presence of recursive relations is immediate from the substitution rule,
\begin{equation}\label{Snrec}
   S^n(b) \, = \, S^{n-1}(S(b)) \, = \, S^{n-1}(c_1 \ldots c_k) 
          \, = \, S^{n-1}(c_1) \ldots S^{n-1}(c_k) \,.
\end{equation}
Note that the concrete expression, that is, the way to pass from the words 
$S^{n-1}(c_1) \ldots S^{n-1}(c_k)$ to the word $S^n(b)$, is $n$-independent. 

\begin{example}
Let us consider the {}Fibonacci substitution $S_F$ which acts as $S_F(a)=ab$,
$S_F(b)=a$. The recursive relations are given by
\begin{equation}\label{fibrec}
   S_F^n(a) \; = \; S_F^{n-1}(a) S_F^{n-1}(b)
   \; , \quad S_F^n(b) \; = \; S_F^{n-1}(a) \, .
\end{equation}
\end{example}
Let us now consider a Sturmian subshift $\Omega_\alpha$. Consider the 
continued fraction expansion of $\alpha$ (for general information on continued 
fractions, see, e.g., \cite{khin,per}),
\begin{equation}\label{continuedfraction}
   \alpha \; = \; \cfrac{1}{a_1+ \cfrac{1}{a_2+ \cfrac{1}{a_3 + \cdots}}}
\end{equation}
with uniquely determined $a_n \in \NN$. The associated rational
approximants $\frac{p_n}{q_n}$ obey 
\begin{equation}
  p_0 \, = \, 0 \, , \; p_1 \, = \, 1 \, , \; 
  p_n \, = \, a_n p_{n-1} + p_{n-2} \, ,
\end{equation}
\begin{equation}\label{qnrec}
  q_0 \, = \, 1 \, , \; q_1 \, = \, a_1 \, , \; 
  q_n \, = \, a_n q_{n-1} + q_{n-2} \, .
\end{equation}
To make things formally similar to the substitution case, define words $s_n$ 
over the alphabet $\mathcal{A}=\{0,1\}$ by 
\begin{equation}\label{sndef}
   s^{}_n \; = \; v_{\alpha,0}(1) \ldots v_{\alpha,0}(q_n)
\end{equation}
and the one-sided sequence $c_\alpha$ by
\begin{equation}\label{cadef}
  c_\alpha \; = \; \lim_{n \rightarrow \infty} s^{}_n \, .
\end{equation}
Of course, $c_\alpha$ is nothing else than $v_{\alpha,0}$ restricted to $\NN$. 
Note that the words $s^{}_n$ have length $q_n$. The equation (\ref{qnrec}) now 
has the following analog on word level,
\begin{equation}\label{snrec}
   s^{}_n \; = \; s_{n-1}^{a_n} s^{}_{n-2} \, .
\end{equation}
The equation holds in this form for $n \ge 3$. The correct initial conditions 
are recovered by 
\begin{equation}
    s^{}_{-1} \, = \, 1 \, , \; s^{}_0    \, = \, 0 \, , \; 
    s^{}_1    \, = \, s_0^{a_1 -1} s^{}_{-1} \, ,
\end{equation}
and with these definitions (\ref{snrec}) also holds for all $n \ge 2$. This
can be proved by using the fact that continued fraction approximants 
$\frac{p_n}{q_n}$ provide the best possible approximation to $\alpha$; see 
\cite{bist} for details. The following useful formula can be deduced from 
(\ref{snrec}).

\begin{proposition}
{}For each $n \ge 2$,
\begin{equation}\label{wunderformel}
   s^{}_n s^{}_{n+1} \; = \; 
   s^{}_{n+1} s_{n-1}^{a_n - 1} s^{}_{n-2} s^{}_{n-1} \, .
\end{equation}
\end{proposition}
\begin{proof} $s^{}_n s^{}_{n+1} = s^{}_n s_n^{a_{n+1}} s^{}_{n-1} = 
s_n^{a_{n+1}} s^{}_n s^{}_{n-1} = 
s_n^{a_{n+1}} s_{n-1}^{a_n} s^{}_{n-2} s^{}_{n-1} = 
s^{}_{n+1} s_{n-1}^{a_n - 1} s^{}_{n-2} s^{}_{n-1}.$
\end{proof}
Again we have found a sequence of words obeying recursive relations tending to 
a limit word $c_\alpha$ such that the hull can be written in the following 
form,
\begin{equation}\label{sturmhullword}
   \Omega_\alpha \; = \; 
   \{ \omega \in \mathcal{A}^\ZZ : {}F_\omega = {}F_{c_\alpha} \} \, .
\end{equation}
The next step is to pass from the recursively generated words to the transfer 
matrices associated to these words. Recall that given a subshift over some 
alphabet $\mathcal{A}$, the potentials $V_\omega$ were generated by replacing 
the symbols in the sequences by real numbers according to some function 
$f : \mathcal{A} \rightarrow \RR$. Thus, one can study local properties of 
the potentials by introducing the following energy-indexed representation of 
words as ${\rm SL}(2,\RR)$ matrices. {}Fix some real energy $E$ and define the 
map $M_E:\mathcal{A} \rightarrow {\rm SL}(2,\RR)$ by
\begin{equation}\label{repr}
   M_E(a) \; = \; 
   \left( \begin{array}{cc} E-f(a) & -1\\1 & 0 \end{array}\right).
\end{equation}
Extend this mapping to $\mathcal{A}^*$ by
\begin{equation}\label{extend}
   M_E(a_1 \ldots a_n) \; = \;  M_E(a_n) \times \cdots \times M_E(a_1) \, .
\end{equation}
{}For the above two classes, the recursions (\ref{Snrec}) and (\ref{snrec}) then 
extend to the associated matrices in a straightforward way. We obtain, for 
example, in the Sturmian case
\begin{equation}\label{Mnrec}
   M^{}_n \; = \; M^{}_{n-2} M_{n-1}^{a_n} \, ,
\end{equation}
where we set $M^{}_n=M_E(s_n)$. Note that due to (\ref{extend}), the order of 
the factors has been reversed. Depending on the explicit form of a substitution 
$S$, we have a similar analog in this case. In principle, we would like to 
apply the matrix trace to these equations and to study the dynamics, that is, 
the limit $n \rightarrow \infty$, for the traces. This is motivated by a simple 
argument which we will discuss in a moment. However, since the trace is not 
multiplicative, this transition is not as straightforward as the transition 
from words to matrices. This can be remedied by using appropriate identities 
to break down powers and by extending the set of underlying variables (from a 
set of size $|\mathcal{A}|$ to a larger, but still finite, set). {}First of all, 
all the powers can be broken down to one just by using the characteristic 
equation of unimodular matrices $M$, that is,
\begin{equation}\label{squareexp}
   M^2 \; = \; {\rm tr}(M) M - I \, .
\end{equation}
Moreover, one can pass to products not having multiple occurrences of factors 
by using the equation (see \cite{kn})
\begin{equation}\label{doubleocc}
   {\rm tr}(M N M O) \; = \; 
   {\rm tr}(M N){\rm tr}(M O) + {\rm tr}(N O) - {\rm tr}(N) {\rm tr}(O) \, .
\end{equation}
This set of remaining possible products, the enlarged alphabet 
$\mathcal{E}$, has cardinality bounded by \cite{kn} (see \cite{ab,abg} 
for improvements)
\begin{equation}
   \sum_{l=1}^{|\mathcal{A}|} \frac{|\mathcal{A}|!}{l (|\mathcal{A}| - l)!} \, .
\end{equation} 
Although one may obtain messy expressions, this method generates polynomial 
expressions for the traces of these products of level $n$ in terms of the 
traces on level $n-1$. In case of a two-letter alphabet $\mathcal{A}=\{a,b\}$ 
one may choose $\mathcal{E}=\{a,b,ab\}$.  

\begin{example}
In the {}Fibonacci case we infer from (\ref{fibrec}) that 
$x_n(E) = {\rm tr}(M_E(S_F^n(a))$, $y_n(E) = {\rm tr}(M_E(S_F^n(b))$, 
$z_n(E) = {\rm tr}(M_E(S_F^n(a)M_E(S_F^n(b))$ obey
\begin{eqnarray*}\label{fibtm}
x_n(E) & = & z_{n-1}(E),\\
y_n(E) & = & x_{n-1}(E),\\
z_n(E) & = & x_{n-1}(E)z_{n-1}(E) - y_{n-1}(E).
\end{eqnarray*}
In this example we do not even need the enlarged alphabet since 
$z_n(E)=x_{n+1}(E)$. Thus, one has the equivalent recursion 
$$
   x_n(E) \; = \; x_{n-1}(E)x_{n-2}(E) - x_{n-3}(E)
$$
involving only the $x$-variables.
\end{example}
Now why should one expect a connection between the trace map and the spectrum 
$\Sigma$? By the combinatorial subshift definitions (\ref{combihull}) and 
(\ref{sturmhullword}) and by the recursions (\ref{Snrec}) and (\ref{snrec}), 
any operator $H_\omega$ is a strong limit of operators $H_n$ with periodic 
potentials where the periods have length $|S^n(a)|$ (resp., $q_n$) and the 
potential values are obtained by applying $f$ to the symbols in $S^n(a)$ 
(resp., $s_n$). The $\omega$-dependence of these operators is solely reflected 
in the locations of the periods, the actual periodic blocks are the same. 
Write $x_n(E)$ for ${\rm tr}(M_E(S^n(a)))$ (resp., ${\rm tr}(M_n)$). By the 
general periodic theory \cite{rs4} we have
\begin{equation}\label{perspec}
   \sigma(H_n) \; = \; \{E : |x_n(E)| \le 2\} \, .
\end{equation}
Thus, the strong approximation gives (see \cite{rs1})
\begin{equation}\label{strongspectra}
   \Sigma \; = \; \sigma(H_\omega) \subseteq \bigcap_{k \in \NN} 
   \overline{\bigcup_{n \ge k} \{E : |x_n(E)| \le 2\} } \, .
\end{equation}
If one can prove that $\bigcup_{n \ge k} \{E : |x_n(E)| \le 2\}$ is closed, 
then one ends up with a nice (and useful!) property of energies in the 
spectrum, namely, boundedness of traces on a subsequence (even with a uniform 
bound),
\begin{equation}\label{spectraprop}
   \Sigma \; \subseteq \; 
  \bigcap_{k \in \NN} \bigcup_{n \ge k} 
  \{E : |x_n(E)| \le 2\} = \{ E : |x_n(E)| \le 2 \; \,
  {\rm for} \; {\rm infinitely} \; {\rm many} \; n\} \, .
\end{equation}

It turns out that the trace map is a good tool to establish such a property. 
In many cases it can even be shown that equality holds in (\ref{spectraprop}). 
Namely, building upon S\"ut\H{o} \cite{s5}, the following has been shown for 
Sturmian models by Bellissard et al.~\cite{bist}.

\begin{theorem}[Trace map characterization of spectra of Sturmian models]
\label{charspecsturm}
Let $(H_\omega)_{\omega \in \Omega}$ be a Sturmian family corresponding to 
$\alpha,\lambda$. Define $C_\lambda = 2 + \sqrt{8 + \lambda^2}$. Then, we have
$$
   \Sigma \; = \; 
   \{ E : |x_n(E)| \le 2 \; \, {\rm for} \; {\rm infinitely} \; 
   {\rm many} \; n\} = \{ E : |x_n(E)| \le C_\lambda \; \forall n\} \, .
$$
\end{theorem}
Under certain assumptions, so-called {\it semi-primitvity} of the trace map 
and the occurrence of a square in the substitution sequence, which were shown 
to be satisfied by many prominent examples including {}Fibonacci, period 
doubling, binary non-Pisot, and Thue-Morse, Bovier and Ghez obtained a trace 
map characterization of the spectrum in the primitive substitution case. We 
refer the reader to their paper \cite{bg1} and also its precursors 
\cite{b2,bbg1} for a precise statement of the results. We note, however, that 
they do not obtain boundedness of orbits for energies from the spectrum in 
their general setting but rather boundedness on a subsequence. In this sense 
the trace maps associated to primitive substitution models exhibit only a 
partial analogy to Theorem \ref{charspecsturm}.

\section{Partitions and uniform results}
In this section we focus on the local structures of sequences in a subshift 
which is induced by a recursively generated infinite word. As we saw above, 
this class comprises subshifts generated by substitution sequences and Sturmian 
words. We exhibit a uniform combinatorial property for the elements in the 
subshift. We then discuss how to employ this uniform combinatorial property to 
obtain uniform spectral properties.

To keep the notation simple we shall focus on Sturmian subshifts. However, the 
reader may easily verify that the ideas we present here apply to substitution 
models equally well. {}Fix some irrational $\alpha \in (0,1)$. Recall the 
description of a Sturmian hull $\Omega_\alpha$ in terms of a subword definition 
(\ref{sturmhullword}). {}From this, one sees that the elements of $\Omega_\alpha$ 
have a uniform combinatorial property, namely, they have the same set of 
factors which is equal to the set of factors of $c_\alpha$. Now, apart from 
its subword structure, $c_\alpha$ has an additional structure. Namely, it is 
built from blocks $s_n$ which obey a recursive relation. {}From (\ref{cadef}) 
and (\ref{snrec}) one can infer that for each $n \in \NN$, $c_\alpha$ can be 
decomposed into blocks of type $s_n$ and $s_{n-1}$. This decomposition is even 
unique. Moreover, in this decomposition, blocks of type $s_{n-1}$ are always 
isolated and blocks of type $s_n$ have multiplicity $a_{n+1}$ or $a_{n+1} + 1$. 
It is now natural to ask whether this structure is inherited by the elements 
of $\Omega_\alpha$. After all,  as was discussed in the previous section, we 
already know that these blocks have useful properties in that the transfer 
matrices associated to them can be analyzed. It turns out that all these 
properties persist when passing from $c_\alpha$ to 
$\omega \in \Omega_\alpha$ \cite{dl1}.

\begin{proposition}\label{partitionlemma}
Let $\omega \in \Omega_\alpha$. Then for every $n \in \NN$, there exists a 
unique decomposition of $\omega$ into blocks of type $s_n$ and $s_{n-1}$. In 
this decomposition, the multiplicity of each occurrence of $s_n$ {\rm (}resp., 
$s_{n-1}${\rm )} is $a_{n+1}$ or $a_{n+1} + 1$ {\rm (}resp., $1${\rm )}.
\end{proposition}
This decomposition is called {\it $n$-partition of $\omega$}\index{partition}. 
The above result indicates that to a certain extent, the members of 
$\Omega_\alpha$ are equally well accessible to the pointwise methods 
introduced above. Indeed, in order to establish spectral properties that 
cannot be studied by pointwise convergence uniformly, one may apply a 
pointwise criterion to each $\omega$ separately. {}From Proposition 
\ref{partitionlemma} we learn that local repetitive structures abound in 
Sturmian potentials. {}For instance, we know that for every $n$, each 
$s_{n-1}$-block in the $n$-partition is followed and preceded by at least 
$a_{n+1}$ copies of $s_n$, respectively. 

Qualitatively, we have the same phenomena in hulls generated by primitive 
substitution hulls, the specific properties, however, depending on the 
concrete substitution given.

\section{Absence of eigenvalues: Locating squares and cubes}
We now turn to results on absence of point spectrum for models generated by 
circle maps and primitive substitutions. With one exception -- the paper by 
Hof et al.~\cite{hks} which employs a criterion in similar spirit using 
palindromes rather than powers (see \cite{b4} for extensions) -- virtually all the 
known results were 
obtained by using variants of the Gordon method. In this section we show how 
the methods and tools presented in the three preceding sections can be combined 
to yield these results.

Let us begin by discriminating between the types of results that have been 
obtained. Given a family $(H_\omega)_{\omega \in \Omega}$ such that the 
underlying dynamical system $(\Omega,T)$ is a strictly ergodic subshift over 
some alphabet $\mathcal{A}$ with unique invariant measure $\mu$, define
$$
   \Omega_c \; = \; 
  \{ \omega \in \Omega : \sigma_{{\rm pp}}(H_\omega) = \emptyset \} \, .
$$
We say that absence of eigenvalues for $(H_\omega)_{\omega \in \Omega}$ is
{\it generic}\index{generic absence of eigenvalues} if $\Omega_c$ is a dense 
$G_\delta$ (i.e., a countable intersection of open sets which is dense in 
$\Omega$), {\it almost sure}\index{almost sure absence of eigenvalues} if 
$\Omega_c$ has full $\mu$-measure, and {\it uniform}\index{uniform absence 
of eigenvalues} if $\Omega_c = \Omega$. Let us recall some general arguments 
that are useful in this context. {}First of all, to establish a generic result 
it is sufficient to exclude eigenvalues for just one $\omega \in \Omega$.

\begin{proposition}\label{genres}
If $\Omega_c$ is non-empty, then it is a dense $G_\delta$.
\end{proposition}
\begin{proof}
Simon has shown that $\Omega_c$ is a $G_\delta$ \cite{s2}. If $\Omega_c$ is 
not empty, then it contains an entire orbit which is dense by minimality.
\end{proof}
Next, by invariance $\Omega_c$ has $\mu$-measure $0$ or $1$. In order to 
establish an almost sure result, it therefore suffices to bound the measure of 
$\Omega_c$ from below by a positive number. Here is a more elaborate version 
of this which is useful in connection with local investigations.

\begin{proposition}\label{asres}
Suppose $G(n)$, $n \in \NN$, are Borel sets such that
\begin{enumerate}
\item $\limsup_{n \rightarrow \infty} G(n) \subseteq \Omega_c$,
\item $\limsup \mu(G(n)) > 0$.
\end{enumerate}
Then, $\mu(\Omega_c) = 1$.
\end{proposition}
\begin{proof}
The assertion follows from
$$
   \mu(\limsup_{n \rightarrow \infty} G(n)) \; \ge \; 
   \limsup_{n \rightarrow \infty} \mu(G(n)) \, ,
$$
which is readily verified.
\end{proof}
As pointed out earlier, one cannot expect a general way to establish uniform 
absence of eigenvalues for the models we consider here by inspecting a set of 
$\omega$'s which is strictly smaller than $\Omega$. To a certain extent, this 
can be understood in view of the discreteness of the potential values and the 
well-known and heavily studied sensitivity of point spectrum with respect to 
rank one perturbations \cite{sw,dms,g2b,s2b}. Thus one is led to consider each 
$\omega$ individually and to apply pointwise methods. 

It seems interesting to note that a study of the eigenvalue problem motivates 
three different viewpoints, namely, topological arguments for generic results, 
measure-theoretical arguments for almost sure results, and combinatorial 
arguments for uniform results.

Let us now combine these general strategies with the Gordon-type criteria from 
Section 4. We will treat generic, almost sure, and uniform results separately. 

\subsection{Generic results}
We immediately deduce as a first application a criterion for generic absence 
of eigenvalues as follows.

\begin{proposition}
Suppose there exists $\omega \in \Omega$ such that $V=V_\omega$ obeys the 
assumption of either Lemma \ref{zweiblock} or Lemma \ref{dreiblock}. Then, 
$\Omega_c$ is a dense $G_\delta$.
\end{proposition}
A single element with Gordon-type symmetries was found in the {}Fibonacci case 
\cite{s5}, in the general Sturmian case \cite{bist,d2}, and for a class of 
substitution models including period doubling and binary non-Pisot \cite{d4b}. 
We also want to remark that the work by Hof et al.~provides a method to prove 
generic absence of eigenvalues by studying palindromic structures in the 
potentials \cite{hks}. However, their method seems to be restricted to sets of 
measure zero \cite{dz,d6} and is thus not able to establish almost sure or 
uniform results. 

\subsection{Almost sure results}
Similarly, one gets a criterion for almost sure absence of eigenvalues as 
follows. We start with the three-block method. Define 

$$
   G(n) \; = \; 
   \{ \omega \in \Omega : V_\omega(k-n) = V_\omega(k) = V_\omega(k+n)\, , 
   \; 1 \le k \le n \} \, .
$$
Obviously, the $G(n)$ are Borel sets since they are finite unions of cylinder 
sets. Combining Lemma \ref{dreiblock} and Proposition \ref{asres}, we obtain 
the following proposition.

\begin{proposition}\label{dreiblockmeth}
Suppose
\begin{equation}\label{limsupgn}
   \limsup_{n \rightarrow \infty} \mu(G(n)) \; > \; 0 \,.
\end{equation}
Then, $\mu(\Omega_c) = 1$.
\end{proposition}
Moreover, $\mu(G(n))$ can be estimated by inspecting frequencies of cubes of 
length $3n$ due to equation (\ref{cylmeasfreq}). An argument that is often 
useful is the following. 

\begin{lemma}\label{viertepotenz}
Suppose $\Omega = \Omega_\psi$ and there is a fourth power $v^4$ occurring in 
$\psi$ such that $|v| = n$. Then, 
\begin{equation}\label{lowermeasbounddrei}
    \mu(G(n)) \;\ge\; n d_\psi(v^4)\, .
\end{equation}
In particular, the assumption of Proposition \ref{dreiblockmeth} is satisfied 
if one can find a constant $B>0$ and a sequence of words $v_k$ with 
$|v_k|=n_k \rightarrow \infty$ as $k \rightarrow \infty$ such that for all 
$k$, $v_k^4 \in {}F_\psi$ and 
\begin{equation}\label{lowerfreqboundvier}
d_\psi (v_k^4) \ge \frac{B}{n_k}.
\end{equation}
\end{lemma}
This criterion has been applied to circle map models as well as substitution 
models. In \cite{dp1}, the following theorem for circle map models has been 
proved.

\begin{theorem}[Delyon-Petritis]
Let $\Omega_{\alpha,\beta}$ be a circle map hull. Suppose that the coefficients 
$a_n$ in the continued fraction expansion of $\alpha$ obey 
\begin{equation}\label{lsange5}
    \limsup_{n \rightarrow \infty} a_n \;\ge\; 5 \, .
\end{equation}
Then for every coupling constant $\lambda$, eigenvalues are almost surely 
absent.
\end{theorem}
The condition (\ref{lsange5}) has been slightly relaxed by Kaminaga in 
\cite{k1}.

\begin{theorem}[Kaminaga]
Let $\Omega_{\alpha,\beta}$ be a circle map hull. Suppose that the coefficients 
$a_n$ in the continued fraction expansion of $\alpha$ obey 
\begin{equation}\label{lsange4}
   \limsup_{n \rightarrow \infty} a_n \;\ge\; 4 \, .
\end{equation}
Then for every coupling constant $\lambda$, eigenvalues are almost surely 
absent.
\end{theorem}
{}For substitution models the criterion in \cite{d4} reads as follows.

\begin{theorem}
Let $u$ be a fixed point of a primitive substitution $S$. Suppose $u$ has a 
fourth power as a factor. Then for the associated operator family 
$(H_\omega)_{\omega \in \Omega}$, we have $\mu(\Omega_c)=1$.
\end{theorem}
The criterion applies in particular to the binary non-Pisot case. Moreover, 
the argument in the proof can be modified to include the period doubling case, 
too (see \cite{d3} for a slightly more direct proof in this case). 

Turning now to the two-block method, we can modify the above steps as 
follows. Define for $n \in \NN$ and $C < \infty$,
$$
   G'(n,C) \; = \; \{ \omega \in \Omega : V_\omega(k) = V_\omega(k+n), 
  \, 1 \le k \le n , \, |{\rm tr}(M_{E,\omega}(n))| \le C \; \forall 
  E \in \Sigma \} \, .
$$
Again the $G'(n,C)$ are finite unions of cylinder sets and hence Borel sets. 
Lemma \ref{zweiblock} and Proposition \ref{asres} now imply the following.

\begin{proposition}\label{zweiblockmeth}
Suppose there exists $C < \infty$ such that
\begin{equation}\label{limsupgnc}
   \limsup_{n \rightarrow \infty} \mu(G'(n,C)) \; > \; 0 \, .
\end{equation}
Then, $\mu(\Omega_c) = 1$.
\end{proposition}
We have the following criterion which is similar to Lemma \ref{viertepotenz}.

\begin{lemma}\label{drittepotenz}
Suppose $\Omega = \Omega_\psi$ and there is a cube $v^3$ occurring in $\psi$ 
such that $|v| = n$ and $|{\rm tr}(M_E(v))| \le C$ for every $E \in \Sigma$. 
Then, 
\begin{equation}\label{lowermeasboundzwei}
   \mu(G'(n,C)) \;\ge\; n d_\psi(v^3) \, .
\end{equation}
In particular, the assumption of Proposition \ref{zweiblockmeth} is satisfied 
if one can find constants $B > 0$ and $C < \infty$ and a sequence of words 
$v_k$ with $|v_k|=n_k \rightarrow \infty$ as $k \rightarrow \infty$ such that 
$v_k^3 \in {}F_\psi$,
\begin{equation}\label{vtracebound}
   |{\rm tr}(M_E(v_k))| \;\le\; C \quad \forall E \in \Sigma \, ,
\end{equation}
and
\begin{equation}\label{lowerfreqbounddrei}
   d_\psi (v_k^3) \;\ge\; \frac{B}{n_k} \, .
\end{equation}
\end{lemma}
These criteria are particularly useful in the Sturmian case since we have a 
uniform trace bound for fixed $\lambda$; compare Theorem \ref{charspecsturm}. 
Moreover, Proposition \ref{partitionlemma} allows one to estimate frequencies 
of $s_n^3$ using (\ref{wunderformel}). Putting this together, one obtains the 
following result.

\begin{theorem}[Kaminaga]
Let $\Omega_\alpha$ be a Sturmian hull. Then for every $\lambda$,
$\mu(\Omega_c) = 1$.
\end{theorem}
The proof given in \cite{k1} does not use this two-block argument but rather 
a more elaborate three-block argument. However, using Proposition 
\ref{partitionlemma} one may also find suitably positioned cubes. We want to 
point out that the use of the two-block argument yields additional information; 
compare Remark \ref{2badv}.

\subsection{Uniform results}
This last theorem can even be strengthened. In fact, uniform absence of 
eigenvalues was recently established for all Sturmian hulls.

\begin{theorem}
Let $\Omega_\alpha$ be a Sturmian hull. Then for every $\lambda$, 
$\Omega_c = \Omega_\alpha$.
\end{theorem}
The proof shows that for all $\lambda, \alpha$, Lemma \ref{zweiblock} is 
applicable to every $V_\omega$ \cite{dl1,dl4}. This is the only uniform 
singular continuity result in this context that is known so far.

Let us now summarize the known results on absence of eigenvalues with 
reference to the respective first proof.

\vspace{5mm}
\begin{center}
\begin{tabular}{|l|c|c|c|}
\hline
Model & generic & almost sure & uniform\\
\hline
Circle maps (every $\lambda,\alpha$, $\beta=\alpha$) & 
            \cite{bist} & \cite{k1} & \cite{dl1,dl4}\\
\hline
Circle maps (every $\lambda,\beta$, a.e. $\alpha$) & \cite{dp1} & 
           \cite{dp1} & open\\
\hline
Circle maps (every $\lambda,\alpha,\beta$) & \cite{hks} & open & open\\
\hline
{}Fibonacci substitution & \cite{s5} & \cite{k1} & \cite{dl1}\\
\hline
Period doubling substitution & \cite{bbg1} & \cite{d3} & open\\
\hline
Binary non-Pisot substitution & \cite{hks} & \cite{d4} & open\\
\hline
Thue-Morse substitution & \cite{dp2} & open & open\\
\hline
Rudin-Shapiro substitution & open & open & open\\
\hline
\end{tabular}
\end{center}
\medskip

\section{Zero-measure spectrum}
In this section we show how the fact that the spectrum has zero Lebesgue 
measure can be proved by using the lower bounds that were established when 
proving absence of eigenvalues. In this sense, the zero-measure property is 
merely a corollary to a proof of absence of eigenvalues if the latter is based 
on the two-block method. In fact, this type of argument virtually recovers all 
the known results on zero-measure spectrum. A more comprehensive discussion of 
this simple but somewhat surprising fact is given in \cite{dl3}.

Recall that $A$ denotes the set of energies where the averaged Lyapunov 
exponent vanishes and that it has zero Lebesgue measure for the operators 
under study. The standard way of proving the zero-measure property is to show 
that the spectrum $\Sigma$ is contained in this set,
\begin{equation}\label{sigmaina}
    \Sigma \,\subseteq\, A \, .
\end{equation}
This can be done with the two-block method as follows. Recall that for every 
$E$, there exists a full measure set $\Omega_E$ such that for 
$\omega \in \Omega_E$, the pointwise Lyapunov exponent $\gamma_{\omega,+}(E)$ 
exists and its value is given by $\gamma(E)$. In fact, it has been shown that 
for $E \in \Sigma$, $\Omega_E = \Omega$ for all substitution models \cite{h} 
and all Sturmian models \cite{dl2}. So to prove (\ref{sigmaina}) it is 
sufficient to show that the two-block method is applicable for some 
$\omega \in \Omega$ since it yields that no solution is decaying at $+\infty$, 
whereas, by the Osceledec result Theorem \ref{osc}, a positive Lyapunov 
exponent would give rise to a solution which is exponentially decaying at 
$+\infty$! This simple argument can be applied to all Sturmian models 
\cite{dl3} (thus recovering the main result from \cite{bist}) and it can also 
be used in a slightly modified form (see, e.g., \cite{d5}) to recover and 
elucidate  results of Bellissard et al.~\cite{bbg1} and Bovier and Ghez 
\cite{bg1} for substitution models including period doubling, Thue-Morse, 
and binary non-Pisot.

The main point here is to emphasize that the occurrence of zero-measure and 
thus Cantor spectrum is natural, given the Kotani result and hierarchical 
structures in the potential, the latter leading to a trace map characterization 
of the spectrum. The additional input of the occurrence of squares is 
unavoidable in the case of Sturmian models and substitution models over a 
two-letter alphabet, and it appears as a natural further assumption in the 
result of Bovier and Ghez for substitution models on larger alphabets.

We summarize the known results on zero-measure spectrum together with 
references to their proofs.

\vspace{5mm}
\begin{center}
\begin{tabular}{|l|c|}
\hline
Model & zero-measure spectrum\\
\hline
Circle maps (every $\lambda,\alpha$, $\beta=\alpha$) & \cite{bist}\\
\hline
Circle maps (every $\lambda,\alpha$, $\beta \not= \alpha$) &open\\
\hline
{}Fibonacci substitution & \cite{s6}\\
\hline
Period doubling substitution & \cite{bbg1}\\
\hline
Binary non-Pisot substitution & \cite{bg1}\\
\hline
Thue-Morse substitution & \cite{bbg1}\\
\hline
Rudin-Shapiro substitution & open\\
\hline
\end{tabular}
\end{center}
\medskip

\section{Quantum dynamics}
How would such spectral properties show up in physical systems (resp., 
observables)? To give a first hint, we are now concerned with the transport 
properties of one-dimensional quasicrystal models, that is, with the long 
time behavior of the unitary groups generated by the operators under study. 
We have seen that the presence of purely singular continuous spectrum seems 
to be the rule for circle map and substitution Hamiltonians. Consequently, we 
discuss recent ideas and results concerning the dynamics of operators with 
purely singular continuous spectra. An analysis of the quantum dynamics 
naturally consists of two parts. On the one hand, one tries to identify 
crucial characteristics of singular continuous spectral measures which enable 
one to obtain bounds on the dynamics. Certain dimensions, such as Hausdorff 
dimension and packing dimension, have proved to be useful in this context. 
On the other hand, one seeks methods to study these dimensions which apply 
to the models of interest. Investigations in these directions are still in 
their early stages. In particular the results for concrete operators are very 
limited as the reader will notice. This, however, should be seen as a 
challenge.

Let us first recall parts of the general theory. Suppose $H$ is a self-adjoint 
operator in $\ell^2(\ZZ^d)$ and $\psi \in \ell^2(\ZZ^d)$ with $\|\psi\| = 1$. 
The spectral measure $\mu_\psi$ of $\psi$ is uniquely defined by
$$ 
   \langle \psi , f(H) \psi \rangle \; = \; \int_\RR f(x) d\mu_\psi (x)
$$
for any measurable function $f$. We are interested in the long time behavior of
$$
  \psi(t) \; = \; e^{-itH} \psi \, .
$$
In the singular continuous regime it is convenient to consider time averaged 
quantities as suggested by the RAGE theorem\index{RAGE theorem} \cite{rs3}, a
common quantity being the Cesaro mean of the moments of order $p$ of the 
position operator for $\psi$, that is,
$$
  \langle \langle | X |^p \rangle \rangle (T) \; = \; 
  \frac{1}{T} \int_0^T \langle\psi(t), | X |^p \psi(t) \rangle dt \, ,
$$
where $| X |^p$ is given by
$$
   | X |^p \; = \; 
   \sum_{n \in \ZZ^d} |n|^p \langle \delta_n, \cdot \rangle \delta_n
$$
with the standard orthonormal basis $(\delta_n)_{n \in \ZZ^d}$ of 
$\ell^2(\ZZ^d)$. Several authors have established lower bounds on 
$\langle \langle | X |^p \rangle \rangle (T)$ in terms of certain continuity 
properties of $\mu_\psi$. Typical bounds provide a power law behavior where 
the power depends on the moment, the continuity (measured by some 
$\alpha \in (0,1)$), and the space dimension in the following way,
\begin{equation}\label{lowerqdbound}
   \langle \langle | X |^p \rangle \rangle (T) \; > \; 
   C_{\psi,p} T^{\frac{p\alpha}{d}}.
\end{equation}
The first type of result in this direction is due to Guarneri \cite{g3} and 
Combes \cite{c2}. It requires uniform $\alpha$-H\"older continuity of 
$\mu_\psi$, that is, $\mu_\psi(I) < C | I |^\alpha$ for every interval $I$ 
with $|I| < 1$, $| \cdot |$ denoting Lebesgue measure. It was extended by Last 
in \cite{l} to measures with non-trivial $\alpha$-continuous component, that 
is, $\mu_\psi$ which are not supported on a set of zero $h^\alpha$ measure, 
where $h^\alpha$ denotes the $\alpha$-dimensional Hausdorff measure
\index{Hausdorff measure}; see also \cite{bcm} and \cite{bt}. A recent result 
by Guarneri and Schulz-Baldes relaxes the requirement that the bound holds for 
all times. They are able to prove a similar bound for a sequence of time 
scales $T_n \rightarrow \infty$ in terms of the packing dimension of 
$\mu_\psi$ which sometimes gives a better exponent; see \cite{gs} for details. 
Similar upper bounds purely in terms of Hausdorff dimensional properties of 
$\mu_\psi$ cannot hold true due to an example in \cite{djls} which shows that 
even a pure point measure $\mu_\psi$ can give rise to a growth rate of 
$\langle \langle |X|^2 \rangle \rangle (T)$ which is arbitrarily close to 
ballistic (compare, however, \cite{s1b}).

In the one-dimensional Schr\"odinger operator case, Jitomirskaya and Last 
developed a beautiful way to study such dimensional properies of spectral 
measures \cite{jl1,jl2,jl3}. Their method is in fact an extension of the 
Gilbert-Pearson theory \cite{gp,g1,kp}. It consists of studying the $\liminf$ 
of 
$$
   \frac{\|\phi_1\|_L^{2 - \alpha}}{\|\phi_2\|_L^{\alpha}}
$$
as $L$ tends to infinity, where $\phi_{1,2}$ are solutions of (\ref{deteve}) 
with ``orthogonal'' boundary conditions at the origin and 
\begin{equation}\label{lnorm}
   \|\phi\|_L \; = \;  
   \left( \sum_{n=1}^{\lfloor L \rfloor} |\phi(n)|^2 + 
   (L - \lfloor L \rfloor) |\phi (\lfloor L \rfloor + 1)|^2 
   \right)^{\frac{1}{2}}.
\end{equation}
This approach has been applied to Sturmian models in \cite{jl1,jl3,d2,dl4}. Those 
works obtain the bound (\ref{lowerqdbound}) for all elements in the 
hull in the 
case where the rotation number has bounded density. It holds for all initial 
states $\psi$ with a positive $\alpha$ which depends on the rotation number 
and the coupling contant. The proof essentially consists of three steps. One 
first linearizes the trace map in order to prove a power-law \textit{upper} 
bound on (\ref{lnorm}), uniformly for all solutions (see \cite{irt,it}). Then 
one proves a similar uniform power-law \textit{lower} bound. Interestingly, in 
this step a Gordon-type argument is the key ingredient. {}Finally, one essentially employs the maximum principle
to infer the desired property of the whole-line 
problem from the analysis of the half-line solution behavior.

\section{Open problems}
In this concluding section we list some open problems. We state explicit 
questions as well as vague directions that seem interesting and important.

\subsection{A constructive proof of Kotani's result}
Theorem \ref{efaprop}, essentially a corollary to Kotani theory, is right at 
the heart of most of the results presented and discussed in this paper. 
Specifically, proofs of absence of absolutely continuous spectrum and 
zero-measure spectrum were possible only after Kotani published this theorem 
in 1989. Its value to the known results therefore cannot be overrated. However,
 we feel that an alternative way of understanding these results would be 
extremely interesting. Kotani's proof of the fact that the set of energies 
where the Lyapunov exponent vanishes has Lebesgue measure zero is indirect and 
inconstructive. It does not give any further information as to why the statement 
of the theorem is true. On the one hand, it would be nice to have an intuitive 
understanding of the very uniform absence of absolutely continuous spectrum. 
This phenomenon, of course, relies heavily on the fact that the potentials take 
only finitely many values. {}For example, circle map potentials with the discontinuous 
characteristic function replaced by a smoother function $f$ seem to exhibit a much 
smoother transition from absolutely continuous spectrum through singular continuous 
spectrum to pure point spectrum as the coupling constant $\lambda$ ranges from $0$ 
through finite values to $\infty$; compare \cite{j2} for the completion of the proof 
of this phenomenon in the case of the almost Mathieu operator (i.e., $f = \cos$) 
and \cite{j1,l2} for the state of the almost Mathieu art as of 1994. 
In particular, absolutely continuous spectrum is present in this case for 
non-zero $\lambda$. On the other hand, it would be interesting to investigate 
the Hausdorff dimension of the spectrum. Kotani's proof does not provide any 
clue how to tackle this problem in the general case. However, in the 1987 
S\"ut\H{o} paper \cite{s5}, a constructive proof of Cantor spectrum was given 
in the {}Fibonacci case at large coupling ($\lambda > 4$). Extending this 
approach, Raymond found in \cite{r} a way to obtain upper bounds on the 
Hausdorff dimension in this case. His method should extend to some rotation 
numbers $\alpha$ other than the golden mean, but already the proof in the 
{}Fibonacci case requires considerable effort. The study of the dimension is 
not only interesting from a purely mathematical perspective. Killip et 
al.~have shown that Raymond's study can be used to establish upper bounds on 
the dynamics in the {}Fibonacci case \cite{kkl}. We want to stress, however, 
that these results are limited to large $\lambda$, one particular $\alpha$, 
and $\beta = \alpha$. This brings us to the next direction of possible future 
research activity. 

\subsection{Spectral dimensions and quantum dynamics}
As discussed in an earlier section, there has been considerable progress in 
the understanding of the dynamics of a Schr\"odinger time evolution in the 
presence of peculiar spectral measures, such as purely singular continuous 
measures with multifractal structures. Such multifractal behavior is expected 
to be present for {}Fibonacci-type operators. Results in this direction for 
concrete operators, however, are extremely limited, and it will be worthwhile 
to pursue such investigations. Let us list some questions and problems.

\begin{enumerate}
\item Develop methods that allow for the investigation of dimensional 
      properties of spectral measures.
\item {}Find sufficient criteria for non-trivial dynamical upper bounds.
\item Study dimensions and dynamics for circle map and substitution 
      Hamiltonians.
\item Is it possible to extend the $\alpha$-continuity result for Sturmian 
      models with bounded density rotation numbers to a larger class or is 
      there in fact a delicate dependence of the transport properties of the 
      operator on the Diophantine properties of the rotation number?
\item A partial answer to the above question could be obtained by an extension 
      of Raymond's bound on the Hausdorff dimension of the spectrum to rotation 
      numbers other than the golden mean since this dimension provides a 
      natural upper bound on the dimension of continuity of the spectral 
      measure. Is it possible, for example, to prove that the Hausdorff 
      dimension of the spectrum is zero for, say, Liouville rotation number 
      and large coupling constant?
\end{enumerate}

\subsection{Complexity and spectral theory}
Consider two-sided sequences over a two-letter alphabet, that is, elements of 
$\mathcal{A}^\ZZ$, where $\mathcal{A} = \{x_1,x_2\} \subseteq \RR$. Given such 
a sequence $s$, we ask what properties of $s$ are crucial in determining the 
spectral type of the associated Schr\"odinger operator, $\Delta + s$ in 
$\ell^2(\ZZ)$, where $\Delta$ denotes the discrete Laplacian. It is clear 
that any spectral type can occur. A possible point of view, namely that 
combinatorial properties of $s$ might discriminate between the several 
spectral types, is discussed in this subsection.

Recall the complexity function $p_s$ which measures the subword complexity of 
some sequence $s$, that is, $p_s(n)$ equals the number of subwords in $s$ 
having length $n$. The condition $p_s(n) \le n$ for some $n$ is equivalent to 
$p_s$ being bounded and $s$ being periodic. Moreover, the aperiodic sequences 
$s$ of minimal complexity (i.e., $p_s(n) = n+1$ for every $n$) are essentially 
just the circle map sequences with $\alpha = \beta$ irrational.
Restricting our attention to sequences which are {\it recurrent} in the sense 
that every subword occurs infinitely often, we can formulate two surprising 
implications.

\begin{enumerate}
\item $p_s$ bounded $\Rightarrow$ $\Delta + s$ has purely absolutely 
      continuous spectrum,
\item $p_s(n) = n+1$ $\Rightarrow$ $\Delta + s$ has purely singular continuous 
      spectrum.
\end{enumerate}
As we saw above, sequences generated by primitive substitutions tend to also 
give rise to operators with purely singular continuous spectrum. This fits 
very nicely into this picture since their combinatorial complexity is also at 
the bottom of the hierarchy: It is always bounded by a linear function 
\cite{q}.

On the other complexity extreme, it seems that we encounter a tendency to 
pure point spectrum. This can be argued as follows. Put some non-trivial 
probability measure $\nu$ on $\mathcal{A}$ (i.e., assign the probability 
$p \in (0,1)$ to one letter and $1-p$ to the other) and consider the product 
measure $\bigotimes_{n \in \ZZ} \nu$ on $\mathcal{A}^\ZZ$. Now it is known 
that almost every sequence with respect to this measure leads to an operator 
with pure point spectrum (by results of Carmona et al.~on localization for 
Bernoulli Anderson models \cite{ckm}). On the other hand, almost every sequence 
$s$ has every word from $\mathcal{A}^*$ as a subword and thus has complexity 
function $p_s(n) = 2^n$. We observe that as complexity is increased, the 
spectral measures become more singular. This raises several questions.

\begin{enumerate}
\item Is there a direct proof of the above observations?  
\item Is there a sharp transition from purely singular continuous spectrum to 
      pure point spectrum?
\item Are there examples of potentials in $\mathcal{A}^\ZZ$ where several 
      spectral types coexist?
\end{enumerate}

\subsection{The eigenvalue problem}
Although there has been considerable effort to study the eigenvalue problem 
for one-dimensional circle map and substitution models, we feel that there is 
still room for improving our understanding of this problem. It is extremely 
puzzling that no counterexample is known to the apparent tendency that the 
spectrum is purely singular continuous, and yet one is not able to prove 
absence of eigenvalues for the entire class of potentials. Instead one is 
currently able to deduce the desired result from certain local symmetries 
such as powers or palindromes. However, these symmetries are not always 
present, as the example of the Rudin-Shapiro substitution shows. In fact, 
this example epitomizes our lack of understanding in that it defies almost 
all known and well-established approaches. Apart from the absence of 
absolutely continuous spectrum, essentially nothing is known about this model. 
Ironically, this again hints at the power of the inconstructive Kotani result. 
This leads us to the following concrete problems.

\begin{enumerate}
\item {}Find an example with non-empty point spectrum or prove absence of 
      eigenvalues for all models.
\item More modestly, find new ways to prove absence of eigenvalues or to prove 
      presence thereof.
\item Concretely, study the Rudin-Shapiro case.
\item Less concretely, try to prove hierarchical structures in the 
      eigenfunctions using hierarchical structures in the potentials (e.g., 
      for Sturmian models or substitution models) which prevent them from 
      being square-summable. In doing so one would also gain important insight 
      into dimensional issues using, for example, the Jitomirskaya-Last theory. 
      The reader may take a look at the very interesting paper \cite{j3} by 
      Jitomirskaya which studies circle map potentials for certain paramter 
      values from this point of view.
\item {}Finally, since the Thue-Morse substitution is among the most prominent 
      primitive substitutions, it would be nice to prove almost sure absence 
      of eigenvalues also in this case.
\end{enumerate}

\subsection{Multi-dimensional models} 
Although the focus of this paper is on one-dimensional models, we also want 
to address the problem of extending some of the known results to analogous 
models in higher dimensions. We hope to have demonstrated that a considerable 
amount of knowledge and results has been accumulated for one-dimensional 
Schr\"odinger operators with circle map or substitution potentials. It is 
somewhat striking that only very little is known for their multi-dimensional 
analogs. This is, of course, due to the fact that most results were proved by 
using the transfer matrix formalism which is a purely one-dimensional concept. 
However, we believe that some results, such as absence of eigenvalues for 
certain models, should extend to higher dimensions, and that these extensions 
should be among the major future objectives in this field.

\bibliographystyle{amsalpha}

\end{document}